# Towards an Integrated Platform for Big Data Analysis

Mahdi Bohlouli [1], Frank Schulz [2], Lefteris Angelis [3], David Pahor [4], Ivona Brandic [5], David Atlan [6], Rosemary Tate [7]

[1] University of Siegen, Germany
[2] SAP Research, Karlsruhe, Germany
[3] Aristotle University, Thessaloniki, Greece
[4] Arctur d.o.o, Slovenia
[5] Technical University of Vienna, Austria
[6] Phenosystems SA, Belgium
[7] University of Sussex, United Kingdom

mbohlouli@informatik.uni-siegen.de, frank.schulz@sap.com, lef@csd.auth.gr, david.pahor@arctur.si, ivona@infosys.tuwien.ac.at, atlan_d@web.de, rosemary@sussex.ac.uk

**Abstract**

The amount of data in the world is expanding rapidly. Every day, huge amounts of data are created by scientific experiments, companies, and end users' activities. These large data sets have been labeled as "Big Data", and their storage, processing and analysis presents a plethora of new challenges to computer science researchers and IT professionals. In addition to efficient data management, additional complexity arises from dealing with semi-structured or unstructured data, and from time critical processing requirements. In order to understand these massive amounts of data, advanced visualization and data exploration techniques are required.

Innovative approaches to these challenges have been developed during recent years, and continue to be a hot topic for research and industry in the future. An investigation of current approaches reveals that usually only one or two aspects are addressed, either in the data management, processing, analysis or visualization. This paper presents the vision of an integrated platform for big data analysis that combines all these aspects. Main benefits of this approach are an enhanced scalability of the whole platform, a better parameterization of algorithms, a more efficient usage of system resources, and an improved usability during the end-to-end data analysis process.

## 1. Introduction

The amount of data produced and processed is expanding at an extreme pace. Two sources of data can be distinguished: human-generated data and machine-generated data, and both present huge challenges for data processing. The big data phenomenon cannot be defined by data volume alone. Additional layers of complexity arise from the speed of data production and the need for short-time or real-time data storage and processing, from heterogeneous data sources, from semi-structured or unstructured data items, and from dealing with incomplete or noisy data due to external factors. All these aspects render the analysis and interpretation of data a highly non-trivial task. It becomes even more challenging when data analysis and decision making needs to be carried out in real time. The information processing capacity of humans is highly limited. One highly cited study showed that only about seven pieces of information can be held in short-term memory [20]. Therefore a suitable technological support is strongly needed in order to present the information in a more accesible form.

Taking these points into consideration, the following definitions have been proposed to capture the notion of big data: "big data is when the size of the data itself becomes part of the problem" (Loukides in [22]) or "data that becomes large enough that it cannot be processed using conventional methods" (Dumbill in [22]). Other authors define big data by the three dimensions of volume, velocity and variety [26].

### 1.1 Big Data Examples

There are numerous domain examples, including web applications, recommender systems for online advertising, financial decision making, medical diagnostics, or the operation of social networks or large IT infrastructures. For instance, Google was processing 20 petabytes ($10^{15}$ bytes) per day in 2008. In 2011, and was able to sort one petabyte of 100-byte-strings in 33 minutes on an 8000 machine cluster. Amazon.com reported peak sales of 158 sold items per second on November 29, 2010, and Walmart retail markets handle more than 1 million of transactions per hour. Nowadays, the amount of data from automated sensors, RFID tags or mobile devices surpasses the human-generated data by far. According to Teradata, a single six-hour flight of a Boeing 737 airplane produces 240 terabyte of sensor data. It was estimated by IBM that currently 2.5 x $10^{18}$ bytes of raw data are created every day by humans or machines [17].

### 1.2 Business Impact

In 2011 for the first time, Gartner Market Research added the term "'Big Data' and Extreme Information Processing and Management" to their annually published hype cycle for emerging technologies. The business value of advanced analytics of huge amounts of data has been widely recognized as a key driver for future business growth. The analyst firm Wikibon published a report that estimates an annual growth of the Big Data market of over 50 % for the next five years, resulting in a market volume of 53 billion US-$ in 2017. Key drivers are the competitive advantage and the increased operational efficiency gained by advanced analytic capabilities.

This paper presents the first ideas and goals of an initiative that was originated as a collaboration of researchers from European universities and companies, aiming to develop a generic, sophisticated, and customisable platform able to extract information from extra-large data sources and streams from the Cloud environment or physically situated resources. Using pattern recognition, statistical and visual analytic techniques, the goal of the integrated platform is to present the information in a helpful form, enhancing decision-making across many domains.

The initiative aims to combine advanced techniques to enable: (a) applications across a wide range of domains; (b) integration of large-scale data from disparate resources and streams; (c) scalability and elasticity on cloud infrastructures; (d) statistical identification and discovery of complex events that would be imperceptible for standard analyses; (e) effective and meaningful decision support, and (f) continuous quality control of results.

In section 2, existing technologies for big data are reviewed. Section 3 discusses some specific use cases and derives requirements for an integrated platform. Section 4 outlines the proposed architecture and key aspects of such a platform. In section 5, related work on end-to-end data analysis platforms is discussed.

**2. Existing Technologies in Data Analysis and Machine Learning**

Decision support systems are nowadays ubiquitous in industrial and research applications, and a large variety of commercial and open source tools and libraries exist. Furthermore, there is a rich theoretical background from various disciplines such as statistics and operations research that lays a solid foundation for decision making systems. The use of statistical and data mining methods has been limited to specific data from specific sources, depending on the application domain.

Notable open source tools include the R project [25] for statistical analysis, the WEKA project [15] for data mining, the KNIME platform [6] for data analytics, and the Apache Mahout [4] project for machine learning on top of the map reduce framework Hadoop. The R statistical language and the Predictive Model Markup Language (PMML) offer the opportunity to combine a wide range of statistical methodologies and models that are able to cooperate for processing massive data from diverse sources and producing output for feeding the decision support systems.

**2.1 R Project and PMML**

R is an open source statistical language, in fact a comprehensive suite of tools providing to the users a vast variety of statistical and graphical techniques for data analysis, and most importantly, the facilities to expand the language by programming new routines, functions and to add new packages. Furthermore, R can be linked with other languages (C, C++) and can be used for advanced massive data analysis.

The Predictive Model Markup Language (PMML) is an XML-based language developed by the Data Mining Group (http://www.dmg.org/) providing ways to represent models related to predictive analytics and data mining. PMML enables the sharing of models between different applications which are otherwise incompatible. The primary advantage of PMML is that the knowledge discovered can be separated from the tool that was used to discover this knowledge, so it provides independence of the knowledge extraction from application, implementation platform and operating system.

**2.2 WEKA**

The Weka workbench [15] is a collection of state-of-the-art machine learning algorithms and data preprocessing tools. It is very flexible for users who can easily apply a large variety of machine learning methods on large datasets. It can support the whole process of data mining, starting from the preparation of data to the statistical evaluation of the models. The workbench includes a wide variety of methods such as regression, classification, clustering, association rule mining, and attribute selection. Furthermore, it supports streamed data processing. The system is open-source software, written in Java and freely available.

**2.3 KNIME**

According to [6], the Konstanz Information Miner (KNIME) is a modular environment, developed as an open source project (http://www.knime.org) which enables easy visual assembly and interactive execution of data pipelines. It is designed as a teaching, research and collaboration platform and provides integration of new algorithms and tools as well as data manipulation or visualization methods in the form of new modules or nodes. Its great advantage is the powerful user interface, offering easy integration of new modules and allowing interactive exploration of analysis results or models. Combined with the other powerful libraries such as the WEKA data mining toolkit and the R statistical language, it provides a platform for complex and massive data analysis tasks. KNIME is continuously maintained and improved through the efforts of a group of scientists and is offered freely for non-profit and academic use.

### 2.4 Apache Mahout

Mahout is an open source software project hosted by the Apache foundation [4]. It provides a machine learning library on top of Hadoop, with the goal to provide machine learning algorithms that are scalable for large amounts of data. The development has been initiated with the paper [8]. Up to now, several dozens of algorithms have been implemented for data clustering, data classification, pattern mining, dimension reduction and some others. All algorithms are written in Java and make use of the Hadoop platform.

### 3. Requirements

This chapter states some requirements on the envisioned integrated platform for big data analysis. Based on the analysis of several use cases from different domains, the following areas have been identified as crucial building blocks.

- Functional requirements
    - Data integration: For addressing problems from real-world application domains, the platform must be capable to access multiple different data sources and to deal with inconsistent or noisy data.
    - Statistical analysis: The analysis of data can be simple (like counting) or complex (like the calculation of a Bayesian network for prediction). The platform must support different types of data analysis, including the calculation of statistical key figures like quantiles or correlation coefficients.
    - Interactive exploration: The platform has to support intuitive visualization for visual analytics and easy interaction with the data.
    - Decision support: In addition to the analysis of data, the platform should also provide mechanisms for domain-specific data interpretations that are valuable for decision making. Manual analysis, semi-automated decision support or fully automated systems should be provided depending on the application area.
- Non-functional requirements
    - Scalability: The platform and its various constituents have to be able to handle huge amounts of data. All components must be designed in such a way that they can be deployed in a distributed computing environment.
    - Near real time: Fast processing is the main requirement of some use cases. The core platform must be able to support near real time processing in combination with selected components.
    - Resource efficiency: While keeping the objectives of throughput and speed, the system resources should be utilized in an efficient way. This has to be achieved by a system management component which is part of the platform.

These requirements can also be categorized with respect to research area:
- Database Technologies: Distributed databases, parallelism, NoSQL approaches.
- Information Systems: Design of an integrated platform with scalability of all components and efficient usage of IT resources, making use of current system architectures (multi-core) and increased availability of main memory.
- Algorithmics: Design of efficient algorithms for external memory, algorithms fitting into the MapReduce paradigm or other parallelization patterns. Streaming algorithms for efficient processing of amounts of data that are so huge that scanning it more than once or a few times only is infeasible, or for processing data that naturally arrives as an event stream.

The main challenge exists in the combination and simultaneous realization of these requirements.

### 4. Solution Appraoch

The proposed platform for data analysis aims to address the requirements given in section 3. The following figure 1 describes its main components.

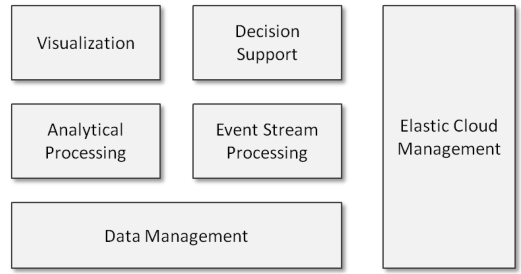

**Fig. 1:** Building blocks

The data management component is responsible for providing access to heterogeneous data sources, data integration and preprocessing. It contains a storage component and interfaces for efficient data retrieval and aggregation. Depending on the requirements, either a batch-oriented analytical processing or a near real-time event stream processing can be performed on the data. In both components, efficient algorithms will provide parallel processing for statistical analysis and complex evaluations. The results can be visualized for human consumption or used as input for a decision support component that provides semi or fully automated solutions in decision problems. While all components have to be scalable in order to cope with large amounts of data, a dedicated management component for elastic clouds will control the infrastructure resources provided to each component for an efficient and balanced operation of the whole system. In addition, this component takes service level objectives into account for ensuring the requested end user experience.

For Big Data manipulation and processing we have concluded that we must use NoSQL databases, which add affordable *horizontal scalability* (scale-out) of storage spreading over nodes, over clusters and eventually over data-centres to *vertical scalability* (scale-up) and enable large data-throughputs - especially write-to-storage. With this decision, we are consciously sacrificing the RDBMS capabilities of orthogonalized data schemes consisting of tables and complex relationships (like joins) and a powerful query language (SQL).

At the same time it is to be stressed that the fundamental differences between today's leading NoSQL solutions are much greater that the differences between different "strains" or products of RDMSs. The NoSQL landscape is filled with disparate and - sometimes - diverging solutions of optimization for Big Data handling that can be complementary only if a unified platform with a common systems' API is implemented. NoSQL databases scale in very different ways, having greatly differing data models and specific mechanisms for data querying. The latter are - on the main part - much more primitive than SQL although attempts are being made recently to bring more structure to querying in certain NoSQL databases - for example by developing SQL-like interfaces, such as Pig, Hive and unQL top of the MapReduce mechanism. Furthermore, there are also significant differences in the type of scaling NoSQL products support. Some of them enable good scaling of the *data-set size*, some grow well in the *volume of concurrent transactions*, some excel only in the brute *speed of data storage read or write,* while others have a hybrid set of the before mentioned scalability capabilities but with significant compromises stemming from this.

The danger of using the wrong NoSQL tool for a specific large data-set processing problem is thus much more pronounced than choosing the "wrong" RDBMS for classic relational processing. Also the implementation, systems integration and programming of some of the NoSQL databases is much more challenging that the incorporation of relational database technologies in applications and middleware due to the young age and documentation scarcity of some of the NoSQL products. Another fact is that not all NoSQL databases are good at (horizontal) distribution over nodes and not all NoSQL databases support effective replication (especially master-to-master) between server clusters. Usually, good scalability paired with excellent node-distribution means the underlying data model is primitive, and vice-verse. A good case in point are graph databases which are very single-node scalable and transaction-throughput efficient but are not optimized for efficient horizontal distribution of processing.

How can we, then, propose our *Platform*, if the NoSQL landscape is so divergent and - partially - exclusive? For the purpose of establishing our *Integrated Platform for Big Data Analysis* we propose the realization of a practical solution for Big Data storage and management that is standardized and formalized as much as possible, while at the same time supports different aspects of strengths of separate NoSQL store-type solutions. On the other hand our combined NoSQL system should be manageable and controllable, so we must limit the number of databases used. Our solution is to use a hybrid middleware NoSQL system for Big Data storage, composed of 3 different databases, each optimized for a specific data model/performance and scalability case. Large data-sets will be thus stored in several different NoSQL (non-relational) databases in the back-end of the *Platform* infrastructure, depending on the type, amount and stream bandwidth of the input data, so that the most optimal database managing system will be used for the appropriate type. It is expected that the different use-case scenarios will provide data and data-streams of events that will demand specialized database processing. Complex querying of sufficiently small sub-sets of data will still be possible with a RDMS.

According to **Brewer's Theorem** - also known as the **CAP Theorem**, distributed computer systems cannot guarantee all of the following:
- **Consistency**, with all network nodes seeing the same data at the same time;
- **Availability**, with a guarantee that every request to the system resources receives a response about whether it was successful or failed;
- **Partition Tolerance**, with the the system continuing to operate despite arbitrary loss of messages between nodes.

Since the CAP theorem states that it is impossible to have both ACID (*Atomic, Consistent, Isolated, Durable*) database consistency and high data availability, we shall use the above described hybrid NoSQL infrastructure that will enable consistency (ACID) for certain usages and high availability for others - depending on the use case. Dissimilar data sources in use cases for the *Platform* will be, therefore, handled by the hybrid storage back-end of the elastic cloud of the *Platform* infrastructure, consisting of an SQL database for smaller data-sets and three specialized types of distributed, multi-nodal NoSQL databases for large data storage, each of them optimized for a certain use scenario (only one database product per store-type):
- **Document Store**, like, for example Apache CouchDB (BigCouch) or MongoDB;
- **Wide Column Store**, such as, for example HBase or Cassandra;
- **Key Value Store**, like, for example MEMBASE, Riak or Redis;

There is a fourth type of NoSQL store - the **Graph Database**, for example InfoGrid, Neo4J or Infinite Graph, which implements flexible graph data models. For our *Platform* we have decided to minimize complexity and have determined that we can cover all major large-data processing with the combination of some or all of the above mentioned store-types, without introducing the added management and programming overhead of Graph Databases.

The infrastructure will have an open services interface based upon a RESTful open data protocol handler positioned between the back-end distributed resources (applications and data) and the service consumers (clients of the use case scenarios) in the elastic cloud. Quality of service is undoubtedly an important factor in today's distributed IT systems with loosely coupled client applications connecting in large numbers to services interfaces of back-end distributed applications in server grids. The *Platform* infrastructure will address this through its specialized framework. The above mentioned NoSQL database systems do not require high-specification servers [29]. In this regard, the systems will not displace current working machines with new resources, but the development of the elastic infrastructure will be accomplished by adding new machines to the current working cluster.

### 5. Conclusion and Outlook

This paper presents the vision of an integrated platform for big data analysis. The vision encompasses the applicability to wide range of domains, the integration of heterogeneous data resources and streams, the efficient usage of computing resources, statistical analysis and machine learning, and an effective and meaningful decision support.

The paper provides a thorough overview of relevant technologies and related work on big data platforms. The key requirements and a high level solution are described. Main benefits of the intended platform are an enhanced scalability of the whole system, a better parameterization of algorithms, a more efficient usage of system resources, and a better usability during the end-to-end data analysis process.